\documentclass[mlmain,onecolumn]{jmlr}

\usepackage{longtable}

\usepackage{booktabs}
\usepackage[load-configurations=version-1]{siunitx} 
\usepackage[dvipsnames]{xcolor} 

\theorembodyfont{\upshape}
\theoremheaderfont{\scshape}
\theorempostheader{:}
\theoremsep{\newline}

\usepackage{soul}

\title[Mixed Monotonicity Reachability Analysis of neural ODE]{Mixed Monotonicity Reachability Analysis of Neural ODE:\titlebreak A Trade-Off Between Tightness and Efficiency}

  \author{\Name{Abdelrahman Sayed Sayed} \Email{abdelrahman.ibrahim@univ-eiffel.fr}\\
  \Name{Pierre-Jean Meyer} \Email{pierre-jean.meyer@univ-eiffel.fr}\\
  \Name{Mohamed Ghazel} \Email{mohamed.ghazel@univ-eiffel.fr}\\
  \addr Univ Gustave Eiffel, COSYS-ESTAS, F-59657 Villeneuve d’Ascq, France}

\begin{document}

\maketitle

\begin{abstract}

Neural ordinary differential equations (neural ODE) are powerful continuous-time machine learning models for depicting the behavior of complex dynamical systems, but their verification remains challenging due to limited reachability analysis tools adapted to them. We propose a novel interval-based reachability method that leverages continuous-time mixed monotonicity techniques for dynamical systems to compute an over-approximation for the neural ODE reachable sets. By exploiting the geometric structure of full initial sets and their boundaries via the homeomorphism property, our approach ensures efficient bound propagation. By embedding neural ODE dynamics into a mixed monotone system, our interval-based reachability approach, implemented in TIRA with single-step, incremental, and boundary-based approaches, provides sound and computationally efficient over-approximations compared with CORA's zonotopes and NNV2.0 star set representations, while trading tightness for efficiency. This trade-off makes our method particularly suited for high-dimensional, real-time, and safety-critical applications. 
Applying mixed monotonicity to neural ODE reachability analysis paves the way for lightweight formal analysis by leveraging the symmetric structure of monotone embeddings and the geometric simplicity of interval boxes, opening new avenues for scalable verification. This novel approach is illustrated on two numerical examples of a spiral system and a fixed-point attractor system modeled as a neural ODE. 

\end{abstract}

\section{Introduction}
\label{sec:intro}

Neural ordinary differential equations (neural ODE) are a recent machine learning model that has been introduced into the machine learning community \citep{rico1992discrete,chen2018neural}, and since then, they have gained prominence in time-series modeling over traditional neural networks \citep{kidger2022neuraldifferentialequations,haber2017stable,oh2025comprehensivereviewneuraldifferential}. As neural ODE constitute a relatively recent technique, fewer verification tools are available for them compared with those for traditional neural networks. In particular, reachability analysis is a useful and important approach for neural ODE, and most of the current available verification tools for neural ODE are based on reachability analysis.

The first work on neural ODE reachability analysis was introduced in \citet{grunbacher2021verification} using Stochastic Lagrangian Reachability, an abstraction-based technique that computes confidence intervals for reachable sets with probabilistic guarantees. This work was later extended to the \emph{GoTube} tool \citep{gruenbacher2022gotube}, enabling computation of neural ODE reachable sets over longer time horizons.
However, both \citet{grunbacher2021verification} and \citet{gruenbacher2022gotube} provide only stochastic bounds on reachable sets, lacking formal guarantees. For deterministic neural ODE reachability analysis, \citet{manzanas2022reachability} have extended the neural network verification tool (\emph{NNV}) \citep{tran2020nnv} by introducing \emph{NNVODE}, a general neural ODE class combining continuous and discrete time layers. \emph{NNV} was later upgraded to \emph{NNV 2.0} \citep{lopez2023nnv}, which offers support for neural ODE. A different form of reachability based formal verification for neural ODE was introduced in \cite{sayed2025bridgingneuraloderesnet}, where a formal relationship between neural ODE and ResNet was introduced to verify safety properties of one model based on the other without running verification tools twice on both models.

Mixed Monotonicity was first introduced in \citet{gouze1994monotone} and then introduced into reachability analysis field in \citet{coogan2015efficient}. The first characterization of continuous-time systems that satisfy mixed monotonicity was investigated in \citet{coogan2016stability,coogan2016mixed}, showing that systems with sign-stable Jacobian matrices satisfy the mixed monotonicity property. This was then generalized for nonlinear systems in \citet{yang2019sufficient}, provided the Jacobian matrices are bounded. Such a result was subsequently applied to the neural network reachability analysis in \citet{meyer2022reachability}, and \citet{coogan2020mixed} further discussed how mixed monotonicity enables efficient reachable set approximation for safety critical dynamical systems.

Some of the tools that consider mixed monotonicity for neural network reachability analysis include \emph{immrax}, a \emph{JAX}-based tool that employs mixed monotonicity to compute interval-based over-approximations of reachable sets for neural networks, utilizing GPU acceleration for computational efficiency \citep{harapanahalli2024immrax}. Additionally, \emph{npinterval} implemented in \emph{numpy}, uses inclusion functions to provide interval bounds on a function's output \citep{harapanahalli2023toolbox}. To the best of our knowledge, no other work has considered mixed monotonicity for the reachability analysis of neural ODE. 

\paragraph{Our Contributions:} We present a novel method for the reachability analysis of neural ODE by adapting existing mixed monotonicity reachability methods for continuous-time dynamical systems \citep{meyer2021interval} to obtain an interval over-approximation of the reachable output set starting from a given input set. We also compare our novel neural ODE mixed monotonicity reachability approach implemented in TIRA \citep{meyer2019hscc} against two well-known reachability analysis toolboxes that can handle neural ODE which are CORA \citep{althoff2015introduction} and NNV2.0 \citep{lopez2023nnv}, in addition to a boundary analysis-based reachability approach inspired by the work of \citet{liang2023safety}.

\section{Preliminaries}
\label{sec:Preliminaries}

We consider the following neural ODE:
\begin{equation}
\dot x(t)=\frac{dx(t)}{dt}=f(x(t)),
\label{eq:nODE}
\end{equation}
with state $x\in\mathbb{R}^n$, initial state $x(0)=u$, and vector field $f:\mathbb{R}^n\rightarrow\mathbb{R}^n$ defined as a finite sequence of classical neural network layers (such as fully connected layers, convolutional layers, activation functions, batch normalization). 
Although the proposed approach in this paper is applicable to any neural ODE with Lipschitz continuous vector field $f$, for simplicity of presentation in the notations relying on the derivatives of $f$, we restrict ourselves to the cases where the vector field is continuously differentiable.

\begin{definition}[neural ODE Reachability]\label{def:nODE Reachability}
Given an initial set $\mathcal{X}_{in}\subseteq\mathbb{R}^n$ and final time $t_f$ for the neural ODE, we define the set of reachable outputs as:
\[\mathcal{R}_{\text{neural ODE}}(\mathcal{X}_{in})=\{y\in \mathbb{R}^n \mid y= \Phi(t_f,u), \ u\in \mathcal{X}_{in}\},\]
\end{definition}

where $\Phi:\mathbb{R}\times\mathbb{R}^n\rightarrow\mathbb{R}^n$ is the solution to the state trajectories of \eqref{eq:nODE} based on the corresponding initial value problem:
$$x(t_f)=\Phi(t_f,x(0))=\Phi(t_f,u).$$
Since we usually cannot compute these output reachable sets exactly, we rely on computing an over-approximation denoted as $\Omega(\mathcal{X}_{in})$ such that $\mathcal{R}_{\text{neural ODE}}(\mathcal{X}_{in})\subseteq \Omega(\mathcal{X}_{in})$.

\subsection{Homeomorphism}
\label{sec:homeomorphism}

A homeomorphism function is a continuous function that preserves topological characteristics, mapping only boundaries to boundaries and interiors to interiors \citep{massey1991basic}. In \citet{liang2023safety}, a homeomorphism is defined as a relational map between two sets that preserves all topological properties between them, as illustrated in \figureref{fig:homeomorphism illustration}. The homeomorphism property was introduced in some reachability analysis works involving a set-boundary reachability method for ordinary differential equations \citep{xue2016reach}, and delay differential equations \citep{xue2020over}. 

\begin{definition}[Homeomorphism]\label{def:Homeomorphism}
For two given sets $\mathcal{X},\mathcal{Y} \subseteq \mathbb{R}^n$, there exists a map $m(.): \mathcal{X}\rightarrow\mathcal{Y}$ which is a homeomorphism w.r.t. $\mathcal{X}$ if it is a continuous bijection and the map inverse $m^{-1}(.): \mathcal{Y}\rightarrow\mathcal{X}$ is also continuous.
\end{definition}

\begin{figure}[htbp]
\floatconts
  {fig:homeomorphism illustration}
  {\caption{Homeomorphism (Right) vs. non-homeomorphism (Left) sets}}
  {\includegraphics[width=0.75\linewidth]{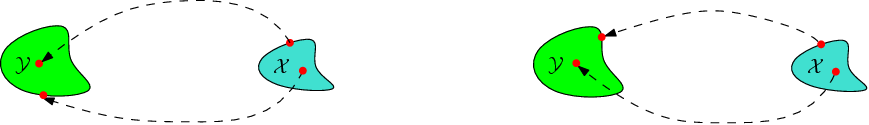}}
\end{figure}

\begin{lemma}[\citet{massey1991basic}]\label{ax:Homeomorphism}
Assuming that the two sets $\mathcal{X},\mathcal{Y} \subseteq \mathbb{R}^n$ are closed and bounded, i.e., compact. For a homeomorphism map $m(.): \mathcal{X}\rightarrow\mathcal{Y}$, $m$ maps the boundaries of the set $\mathcal{X}$ to the boundaries of the set $\mathcal{Y}$, and the interior of the set $\mathcal{X}$ to the interior of the set $\mathcal{Y}$.
\end{lemma}

Neural ODE are naturally invertible \citep{liang2023safety}, which means that they are able to reconstruct their inputs from their outputs, i.e., they correspond to continuous reversible maps. Based on \lemmaref{ax:Homeomorphism}, and the fact that the two sets $\mathcal{X},\mathcal{Y}$ are compact, the neural ODE exhibit the homeomorphism property, which allows us to over-approximate the neural ODE reachable set from the boundaries of its initial set instead of over-approximating the neural ODE reachable set from the full initial set. Based on that, we propose a mixed monotonicity set boundary reachability method for neural ODE, and their corresponding computation procedure is presented in \sectionref{sec:boundary reachability} and \algorithmref{alg:boundary-mm-nODE}.

\subsection{Continuous time Mixed Monotonicity}
\label{sec:CT Mixed monotonicity}

Mixed monotonicity is a property of dynamical systems that allows for efficient computation of reachable sets by decomposing the system's vector field into components that are monotonically increasing and decreasing in their arguments \citep{coogan2020mixed}. Such decomposition allows the embedding of the original system into a higher-dimensional monotone system, where the bounds can be propagated forward to over-approximate the reachable sets \citep{meyer2021interval}.

\begin{definition}[Neural ODE Mixed Monotonicity]\label{def:Mixed-Monotonicity}
The neural ODE \eqref{eq:nODE} is mixed monotone if there exists a decomposition function $g: \mathbb{R}^n \times \mathbb{R}^n \to \mathbb{R}^n$ such that $g$ is increasing in its first argument, $g$ is decreasing in its second argument based on Definition II.1 from \citet{angeli2003monotone}, and $f$ is embedded in the diagonal of $g$: $g(x,x)=f(x)$.\footnote{The enthusiastic reader is encouraged to refer to \appendixref{apd:nODE Mixed Monotonicity} for more details.}
\end{definition}

The mixed monotonicity property is satisfied on any system that is Lipschitz continuous. Lipschitz continuity ensures that the system's vector field has a bounded Jacobian matrix, which is the only condition required to construct such a decomposition function using the continuous-time mixed monotonicity methods from \citep{meyer2021interval}. 

In this paper, we leverage an existing tool based on mixed monotonicity and named TIRA \citep{meyer2019hscc} for neural ODE. Notably, mixed monotonicity is an interval-based approach, relying on hyperrectangular (box) bounds for computational simplicity while ensuring soundness \citep{meyer2021interval}, compared with other approaches focusing on other set representations, such as zonotopes used in CORA \citep{althoff2015introduction}, and star sets in NNV 2.0 \citep{lopez2023nnv}.

\section{Boundary analysis Mixed Monotonicity Reachability for neural ODE}
\label{sec:bound-analysis}

In this section, we present our approach for computing reachable set over-approximations for neural ODE using mixed monotonicity, enhanced with boundary analysis to leverage the homeomorphism property introduced in \sectionref{sec:homeomorphism} (\lemmaref{ax:Homeomorphism}). We use Jacobian bounds based on the neural ODE reachable tube $\mathcal{R}^{\text{tube}}_{\text{neural ODE}}$. These bounds enable two continuous-time mixed monotonicity methods from \citep{meyer2021interval} which are continuous-time mixed monotonicity and sampled-data mixed monotonicity adapted to neural ODE using three different approaches for reachable set over-approximation: single-step, incremental, and boundary-based. These approaches, illustrated in \figureref{fig:Architecture of the approaches}, are evaluated on the spiral and FPA systems.

\begin{figure}[htbp]
\floatconts
  {fig:Architecture of the approaches}
  {\caption{Illustration of the steps for reachability analysis of neural ODE using different tools, methods and mixed monotonicity approaches}}
  {\includegraphics[width=0.75\linewidth]{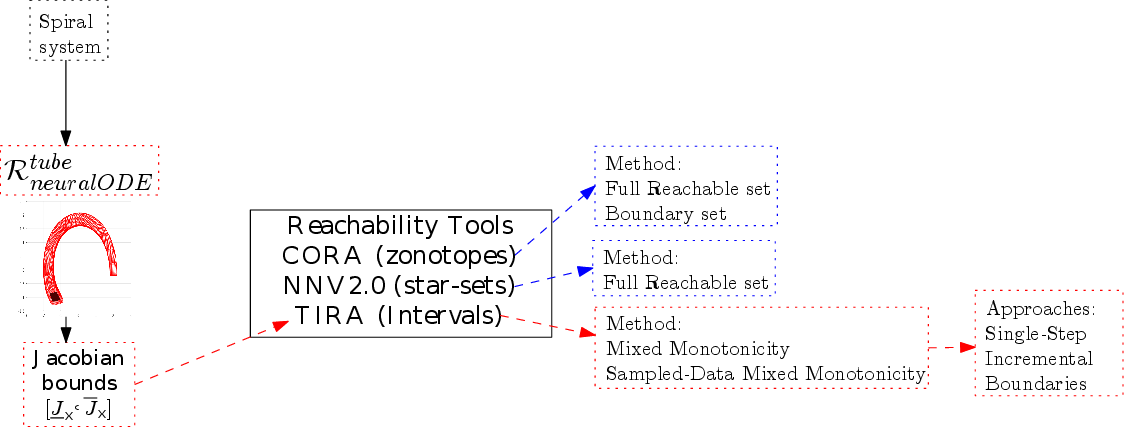}}
\end{figure}

\subsection{Computing the Jacobian bounds}
\label{sec:Jacobian Bounds}

To apply mixed monotonicity, we first need to satisfy the only required condition which is to bound the Jacobian matrix of the neural ODE vector field $f(x)$, ensuring the existence of a sign-stable decomposition \citep{yang2019sufficient}. Using CORA or NNV2.0 \citep{althoff2015introduction,lopez2023nnv}, we compute the neural ODE reachable tube $\mathcal{R}^{\text{tube}}_{\text{neural ODE}}$ over the time interval $[0,1]$ for the spiral system and $[0,2]$ for the FPA system, as a sequence of zonotopes, where each zonotope corresponds to an intermediate time range\footnote{We performed the test while using a longer time range than the ones used in our examples and the resulting Jacobian matrices made the over-approximations not as tight as using only the Jacobian matrices based on the reachable tube of our examples time interval.}. We then compute the union of all the reachable sets in $\mathcal{R}^{\text{tube}}_{\text{neural ODE}}$ to approximate the region containing all neural ODE trajectories over the time interval. From this union, we extract the minimum and maximum values along each dimension across all the reachable sets corresponding to the lower and upper bounds of the reachable tube. Then based on these bounds, we use interval arithmetic \citep{jaulin2001interval} as it is a simple and computationally efficient way to derive bounds on the Jacobian matrices, which will be used as the foundation for the over-approximation methods of continuous-time mixed monotonicity\footnote{For further details about bounding the Jacobian matrices and its usage in mixed monotonicity, please refer to \appendixref{apd:Jacobian bounds,apd:nODE Mixed Monotonicity}}~\citep{meyer2021interval}.

\subsection{Single-Step Reachability analysis}
\label{sec:single-step reachability}

In the single-step approach, we compute the reachable set over-approximation directly from the initial time to the final time $t_f$ without intermediate subdivisions. For the neural ODE~\eqref{eq:nODE} with initial set $\mathcal{X}_{in}$, we embed the system into a mixed monotone form using the decomposition function $g(x, \hat{x})$ derived based on Jacobian bounds. The embedded system is solved over the full time horizon $[0, t_f]$, with $t_f = 1$ for the spiral system and $t_f = 2$ for the FPA system, propagating the interval bounds $[\underline{x}(0), \overline{x}(0)]$ from $\mathcal{X}_{in}$ to obtain the interval over-approximation $\Omega(\mathcal{X}_{in}) = [\underline{x}(t_f), \overline{x}(t_f)]$.

This approach is computationally efficient, as it involves a single integration of the embedded monotone system, providing a direct interval over-approximation of the reachable set at the final time. It is particularly well-suited for systems where achieving a balance between simplicity and soundness is more important than obtaining the tightest bounds.

\subsection{Incremental Reachability analysis}
\label{sec:incremental reachability}

The incremental approach refines the single-step method by dividing the time horizon into smaller steps, computing intermediate reachable set over-approximations to potentially reduce conservatism. For each step, we apply the mixed monotonicity embedding and propagate the bounds sequentially, using the output of one step as the input for the next one.

For the FPA system, we use a step size of $0.05$, resulting in $40$ incremental reachable set over-approximations, i.e., $\frac{t_f}{\text{step size}}$. For the spiral system, a finer step size of $0.01$ is considered, resulting in $100$ over-approximations. 
This method can yield tighter over-approximations than the single-step method, especially for systems with varying dynamics, but at the cost of increased computational time due to repeated numerical integrations of the embedded monotone system.

\subsection{Boundary Reachability analysis}
\label{sec:boundary reachability}

To take advantage of the homeomorphism property of neural ODE, we extend the single-step approach to compute the over-approximation of the reachable set from the boundaries of the initial set only. This can reduce the computation by focusing on the boundary of $\mathcal{X}_{in}$ rather than the entire set, as the interior reachable states are enclosed by the boundary evolution \citep{liang2023safety,xue2020over,xue2016reach}.

For the spiral 2-dimensional system, we compute the reachable sets from $2\times n = 4$ boundaries (the faces of the \textcolor{blue}{dashed-zonotope} in \figureref{fig:spiral_CORA_with_Black_Points} of \appendixref{apd:Detailed Numerical Results}). For the FPA 5-dimensional system, this involves $2 \times 5 = 10$ boundaries (the intervals with the \textcolor{blue}{blue} and \textcolor{red}{red} boundary points in \figureref{fig:FPA_10_Boundary_points_TIRA} of \appendixref{apd:Detailed Numerical Results}). For each boundary, we apply the mixed monotonicity embedding and integrate the embedded system over the full time horizon $[0,t_f]$. We then take the interval hull of the union of the resulting interval over-approximations to form the final over-approximation $\Omega(\mathcal{X}_{in})$. This approach remains sound by enclosing all reachable sets within the boundary evolution, while leveraging the homeomorphism property of the neural ODE, ensuring that the invertible flow map preserves the topological structure of $\mathcal{X}_{in}$. This approach offers computational efficiency for both low and high dimensional systems, as it scales linearly with the state dimension.

\section{Numerical illustration}
\label{sec:experiments}

In this section, two commonly used neural ODE academic examples \citep{chen2018neural,musau2018continuous}, described in detail in \appendixref{apd:Spiral System description,apd:FPA System description}, are used to demonstrate our neural ODE mixed monotonicity reachability approaches. The spiral 2-dimensional example results at $t=1$ are illustrated in \tableref{tab:Spiral_results} and \figureref{fig:Spiral_Comparison_TIRA_CORA_x1_x2_t_1}, while the FPA 5-dimensional example results at $t=2$ are presented in \tableref{tab:FPA_results} and \figureref{fig:FPA_Comparison_TIRA_CORA_Combined Figs_t_2}, using three subfigures to project the 5-dimensional results into two dimensions for visual comparison with TIRA’s interval boxes.

Using the Jacobian bounds computed in \appendixref{apd:Jacobian bounds}, we satisfy the only condition required to compute the over-approximation of the neural ODE mixed monotonicity based reachability approaches, as discussed in Section~\ref{sec:bound-analysis}.

In the spiral example, as shown in \figureref{fig:Spiral_Comparison_TIRA_CORA_x1_x2_t_1}, the \textcolor{Blue}{single-step} and \textcolor{SkyBlue}{incremental} mixed monotonicity approaches gives identical interval over-approximations, but the \textcolor{SkyBlue}{incremental} approach requires approximately \texttt{96 times} more computational time than the \textcolor{Blue}{single-step} approach. Similarly, the \textcolor{OrangeRed}{single-step} and \textcolor{ForestGreen}{dashed boundary}-based sampled-data mixed monotonicity approaches yield equivalent over-approximations, where the \textcolor{ForestGreen}{boundary}-based method is approximately \texttt{5 times} slower than \textcolor{OrangeRed}{single-step} due to computing reachable sets across the four boundaries of the 2D initial set. Comparing across tools, CORA’s zonotopes and NNV2.0 \textcolor{Tan}{star set} achieve tighter over-approximations than TIRA’s approaches, with CORA’s \textcolor{Blue}{dashed-boundaries} zonotope being the tightest. However, CORA’s computational time is approximately \texttt{25 times} greater, and NNV2.0 is approximately \texttt{6 times} greater than TIRA's \textcolor{Blue}{single-step} mixed monotonicity approach, highlighting TIRA’s efficiency for simpler interval-based representations.

\begin{figure}[htbp]
\floatconts
  {fig:Spiral_Comparison_TIRA_CORA_x1_x2_t_1}
  {\caption{Spiral Comparison between TIRA vs. CORA vs. NNV2.0 at $t=1$}}
  {\includegraphics[width=0.5\linewidth]{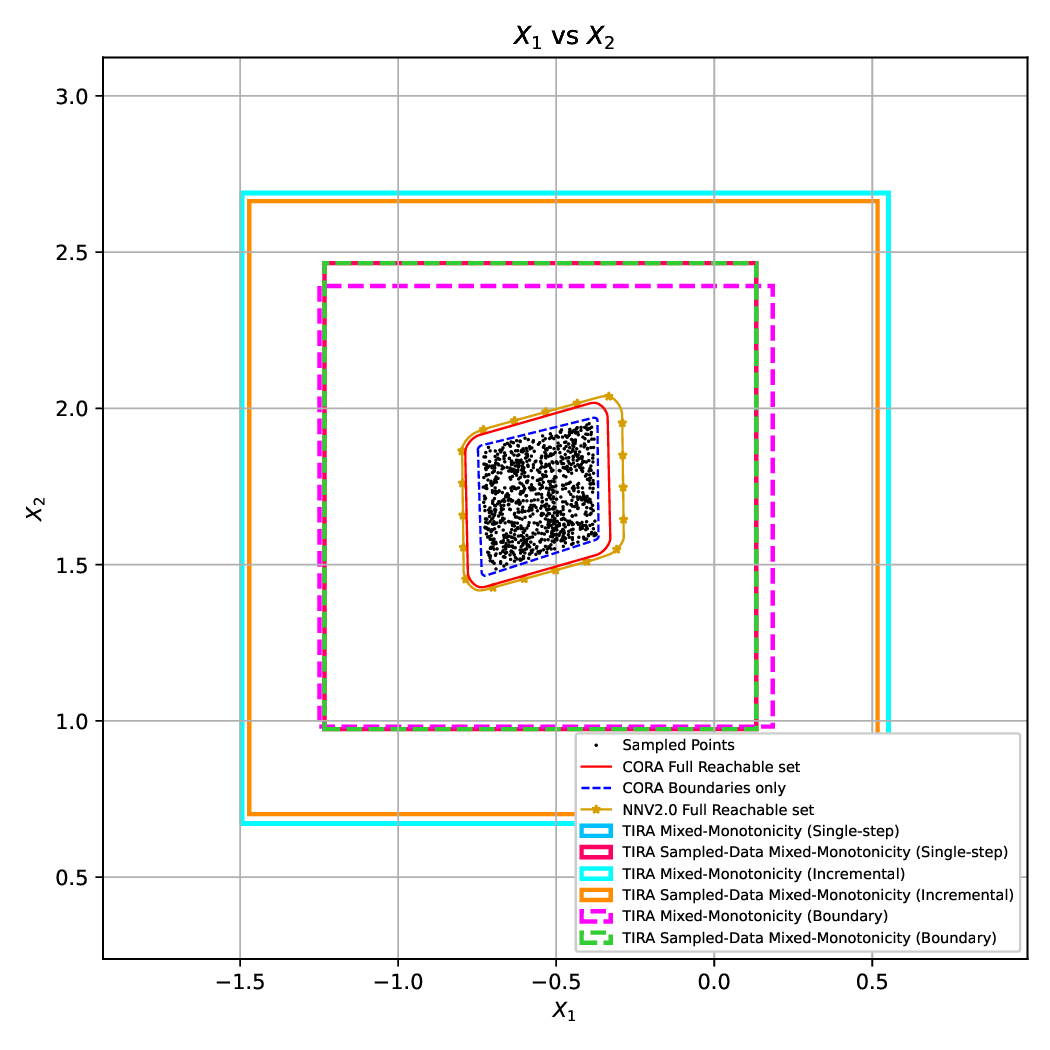}}
\end{figure}

In contrast, for the FPA example illustrated in \figureref{fig:FPA_Comparison_TIRA_CORA_Combined Figs_t_2}, the \textcolor{Blue}{single-step}, \textcolor{SkyBlue}{incremental} and \textcolor{violet}{dashed-boundary}-based mixed monotonicity yield identical over-approximation, but the computational time of the \textcolor{SkyBlue}{incremental} approach is approximately \texttt{31 times} greater than that of the \textcolor{Blue}{single-step} approach, while the \textcolor{violet}{dashed-boundary} approach is approximately \texttt{9 times} greater, making \textcolor{Blue}{single-step} the most efficient choice due to equivalent results.
Additionally, the sampled-data mixed monotonicity approach produces less tight over-approximations than continuous-time mixed monotonicity at a significantly higher computational cost. Across tools, CORA's zonotopes consistently outperform NNV2.0 \textcolor{Tan}{star sets} in tightness with comparable computation times, except for CORA’s \textcolor{blue}{dashed-boundaries} zonotope which achieve the tightest over-approximations overall. However, TIRA’s \textcolor{Blue}{single-step} mixed monotonicity remains the fastest method, with CORA requiring approximately \texttt{131 times} more computation time than TIRA’s \textcolor{Blue}{single-step} mixed monotonicity approach.

For both examples, we used three different set representations to illustrate their strengths and weaknesses. Zonotopes have a more flexible representation, allowing them to model diverse shapes and fit tightly to any set \citep{girard2005reachability}. Thus, for both examples, it is evident that CORA's zonotopes provide the tightest over-approximations compared with NNV2.0 star sets and TIRA's interval boxes, which are very simple rectangular box shapes, but very constrained as well\footnote{Interval boxes cannot be rotated or have their corners truncated.}. In terms of complexity, the simple rectangular box shapes of intervals are easier to compute, resulting in shorter computational times for TIRA compared to CORA and NNV2.0 as illustrated in \tableref{tab:Spiral_and_FPA_timings_results}, and this is due to the fact that zonotopes require processing and storing more data, leading to longer computational times.


\begin{figure}[htbp]
\floatconts
  {fig:FPA_Comparison_TIRA_CORA_Combined Figs_t_2}
  {\caption{FPA Comparison between TIRA vs. CORA vs. NNV2.0 at $t=2$}}
  {%
    \subfigure[$x_1 - x_2$]{\label{fig:FPA_Combined_x1-x2}%
      \includegraphics[width=0.4\linewidth]{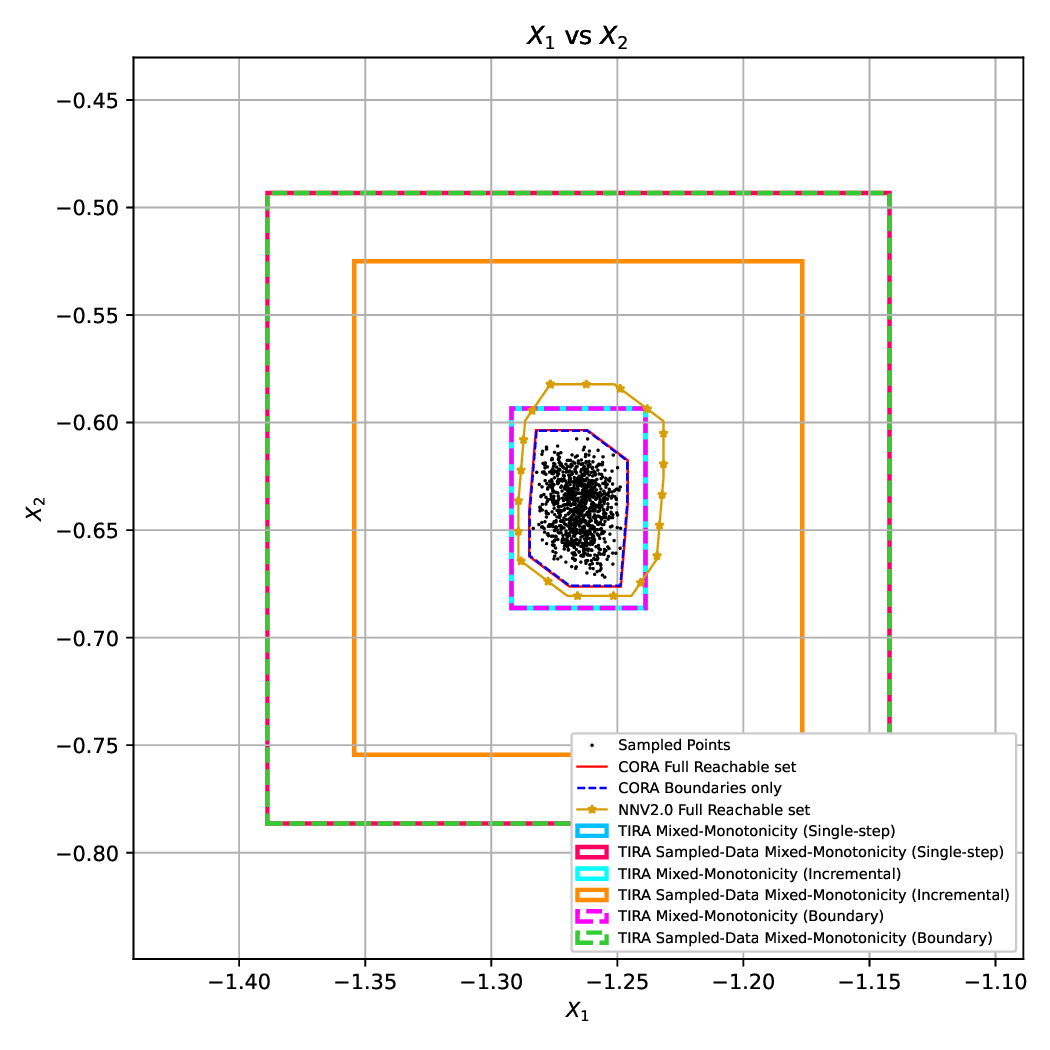}}%
    \qquad
    \subfigure[$x_3 - x_4$]{\label{fig:FPA_Combined_x3-x4}%
      \includegraphics[width=0.4\linewidth]{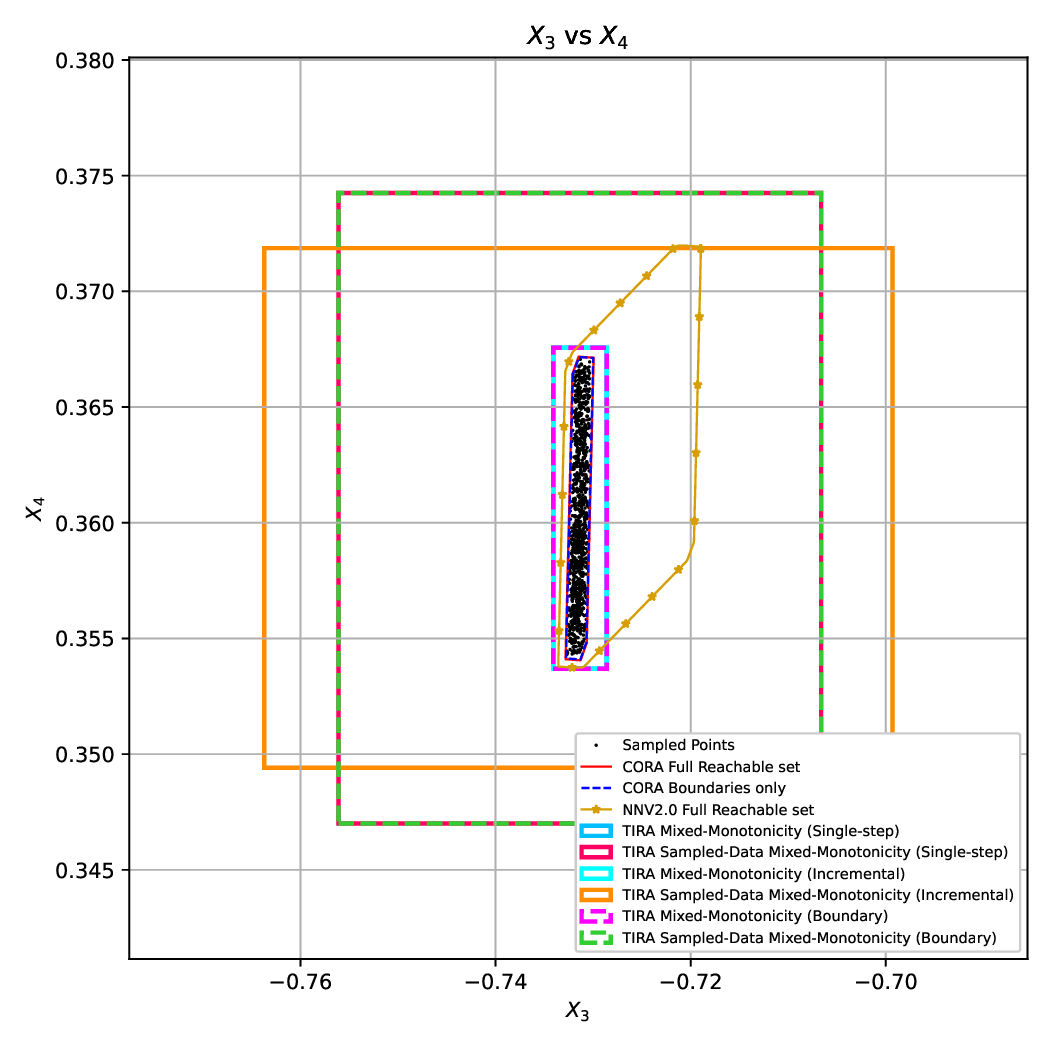}}
    \qquad
    \subfigure[$x_4 - x_5$]{\label{fig:FPA_Combined_x4-x5}%
      \includegraphics[width=0.4\linewidth]{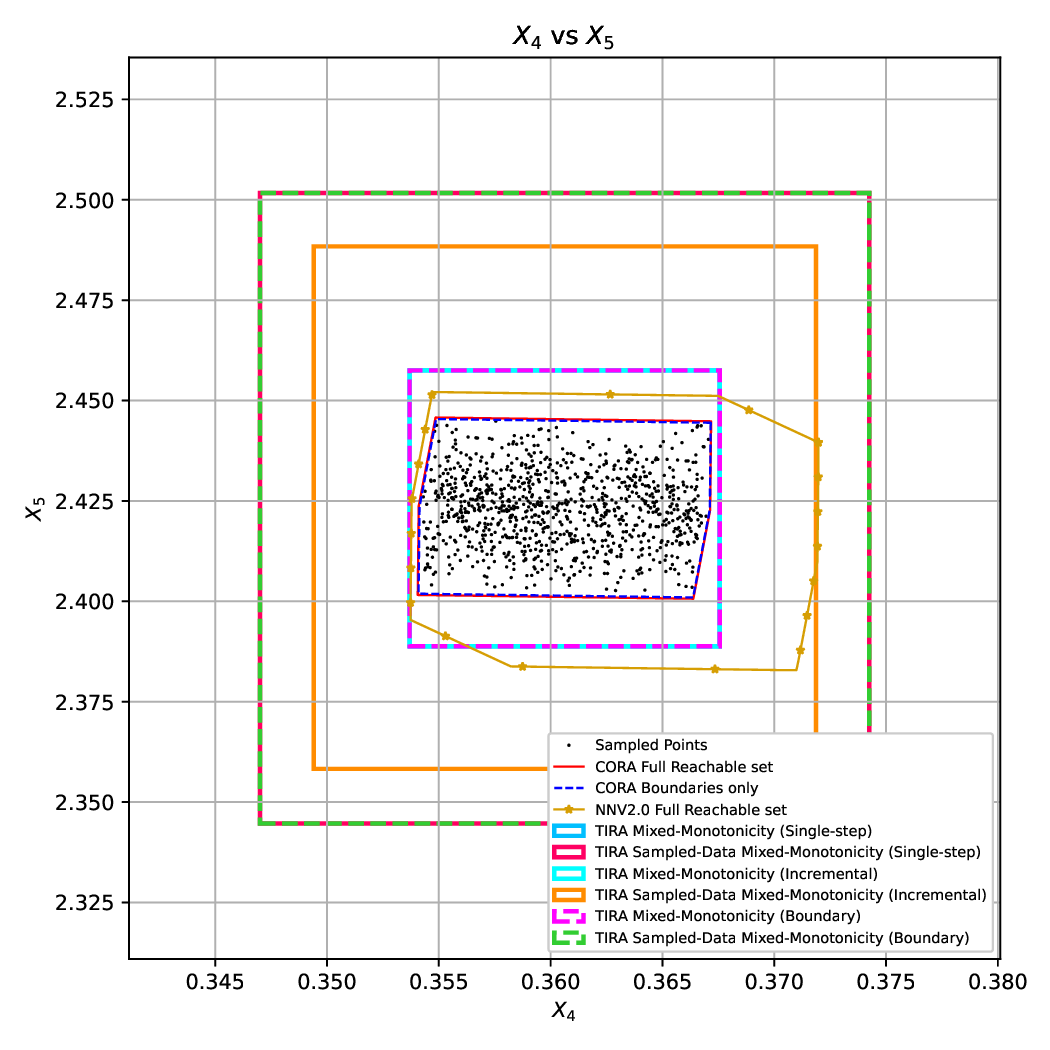}}
  }
\end{figure}

\begin{table}[htbp]
\floatconts
  {tab:Spiral_and_FPA_timings_results}
  {\caption{Spiral and FPA computation times (in seconds) with CORA, NNV2.0, and TIRA}}%
  {%
    \begin{tabular}{|l|l|l|}
    \hline
    \abovestrut{2.2ex}\bfseries Methods & \bfseries Spiral @ $t=1$ & \bfseries FPA @ $t=2$\\\hline
    \abovestrut{2.2ex} CORA Full Reachable Set& 19.64 & 13.22\\
    CORA Boundaries only & 70.83 & 109.1\\
    NNV2.0 Full Reachable Set& 17.25 & 11.98\\
    TIRA (single-step) Mixed-Monotonicity & 0.66 & 0.83\\
    TIRA (single-step) Sampled-Data Mixed-Monotonicity & 0.95 & 1.34 \\
    TIRA (incremental) Mixed-Monotonicity & 63.13 & 25.41\\
    TIRA (incremental) Sampled-Data Mixed-Monotonicity & 111.16 & 48.06 \\
    TIRA (Boundary) Mixed-Monotonicity & 2.84 & 7.06 \\
    \belowstrut{0.2ex}TIRA (Boundary) Sampled-Data Mixed-Monotonicity & 4.35 & 12.76 \\\hline
    \end{tabular}
  }
\end{table}

\section{Conclusions and Future Work}
\label{sec:conc-discuss}

In this paper, we propose an interval-based reachability method for neural ODE, leveraging mixed monotonicity to compute over-approximations of the reachable sets. By adapting continuous-time mixed monotonicity techniques \citep{meyer2021interval}, our approach efficiently computes interval-based over-approximations, both from the full initial input set and its boundaries solely, exploiting the homeomorphism property to reduce computational costs. This novel method, implemented in TIRA, provides a lightweight alternative for the reachability analysis of neural ODE, offering lower complexity and faster computation times compared with more flexible set representations such as zonotopes in CORA and star sets in NNV2.0. 

Through numerical illustrations on the spiral and FPA examples, we demonstrate that our single-step, incremental, and boundary-based variants yield sound over-approximations, albeit at the cost of tightness. Ultimately, the choice between our interval-based methods and zonotope or star set based tools depends on the trade-off between over-approximation tightness and computational efficiency, making our approach particularly suitable for high-dimensional and real-time safety verification scenarios.

In future work, we plan to extend the boundary-based reachability approach by integrating the incremental reachability analysis method to compute over-approximations of the boundaries of the input set over finer incremental steps. We also aim to explore partitioning the initial input set into smaller subsets, performing reachability analysis on each subset using our interval-based method, and subsequently taking the union of all resulting over-approximations. In addition, we intend to incorporate this framework into a full verifier to check safety properties in neural ODE. Finally, evaluating our interval-based reachability method in real-world applications, such as neural ODE-based control systems, typically real-time obstacle detection, will validate the practical utility of the method and guide possible refinements for safety-critical domains.

\acks{This project has received funding from the European Union’s Horizon 2020 research and innovation programme under the Marie Skłodowska-Curie COFUND grant agreement no. 101034248.}

\bibliography{pmlr-sample}

\clearpage
\appendix

\section{System description}\label{apd:System description}

\textbf{Experiments Settings:} All the experiments\footnote{Code available in the following repository:\\ \url{https://github.com/ab-sayed/Mixed-Monotonicity-Reachability-Analysis-of-neural-ODE}} herein are run on MATLAB 2024b with the Continuous Reachability Analyzer (CORA) version 2025.1.1, the Toolbox for Interval Reachability Analysis (TIRA) version 2, and the Neural Network Verification Software Tool (NNV2.0) on an Intel (R) Core (TM) i5-1145G7 CPU@2.60 GHz and 32 GB of RAM. 

\subsection{Spiral}\label{apd:Spiral System description}

The spiral system is a 2-dimensional nonlinear dynamical system modeled as a neural ODE \citep{chen2018neural}, characterized by dynamics that produce spiral trajectories in the state space, making it a valuable benchmark for studying complex, non-convergent behaviors in continuous-time systems, we consider here the following 2-dimensional neural ODE with the following dynamics
$$\dot{x} = f(x) = W_2 \tanh(W_1 x + b_1) + b_2,$$
where $x \in \mathbb{R}^2$ is the state vector, $W_1 \in \mathbb{R}^{10 \times 2}$ is the weight matrix of the first layer, $b_1 \in \mathbb{R}^{10}$ is the bias vector of the first layer, $W_2 \in \mathbb{R}^{2 \times 10}$ is the weight matrix of the second layer, $b_2 \in \mathbb{R}^2$ is the bias vector of the second layer, and $\tanh(\cdot)$ is the hyperbolic tangent activation function applied element-wise to the vector $W_1 x + b_1 \in \mathbb{R}^{10}$. The exact values of the weight matrices and bias vectors are defined within the Matlab function \emph{spiral\_non.m}.

\subsection{FPA}\label{apd:FPA System description}

The FPA system is a 5-dimensional nonlinear dynamical system with dynamics that converge to a fixed point (an equilibrium state) under certain conditions \citep{beer1995dynamics}, and the fixed-point aspect makes it a useful model for studying convergence and stability, which are important in safety-critical applications where the system must not diverge or enter unsafe states.
As in the proposed benchmark in \citet{musau2018continuous}, we consider here the following $5$-dimensional neural ODE approximating the FPA dynamics:
$$\dot{x} = f(x) = \tau x + W \text{tanh}(x),$$
where $x \in \mathbb{R}^5$ is the state vector, $\tau=-10^{-6}$ is a time constant for the neurons, $W \in \mathbb{R}^{5\times5}$ is a composite weight matrix defined as
$W=\begin{pmatrix}0_{2\times 2} & A\\0_{3\times 2} & BA\end{pmatrix}$ \\
with 
$A=\begin{pmatrix}-1.20327 & -0.07202 & -0.93635\\1.18810 & -1.50015 & 0.93519\end{pmatrix}$, and
$B=\begin{pmatrix}1.21464 & -0.10502\\ 0.12023 & 0.19387\\ -1.36695 & 0.12201\end{pmatrix}$.\\
Here, $\text{tanh}(x)$ is the hyperbolic tangent activation function applied element-wise to the state vector $x$. 

\section{Jacobian Bounds}\label{apd:Jacobian bounds}

Considering the FPA system in \appendixref{apd:FPA System description}, the Jacobian  of $f(x)$ is:
$$J_x(x) = \frac{\partial f}{\partial x}(x) = \tau I_5 + W \cdot \text{diag}(1 - \tanh^2(x_i)) = \tau I_5 + W \cdot \text{diag}(\text{sech}^2(x_i)),$$
where $I_5$ is a $5\times5$ identity matrix, and $\text{diag}(1 - \tanh^2(x_i))$ is a diagonal matrix with entries $1 - \tanh^2(x_i)$ for $i = 1, \ldots, 5$, since the derivative of $\tanh(x_i)$ is $1 - \tanh^2(x_i)$. So, the goal is to find interval bounds $[\underline{J}_x, \overline{J}_x]$ such that $J_x(x) \in [\underline{J}_x, \overline{J}_x]$ for all $x \in [\underline{x}, \overline{x}]$.

It is worth noting here that interval arithmetic can be used to directly compute the range of $J_x(x)$ by evaluating the expression over the interval $[\underline{x}, \overline{x}]$. This involves bounding the nonlinear term $1 - \tanh^2(x_i)$ and combining it with the linear terms, as follows:

\begin{enumerate}
    \item For each $i\in\{1,\dots,5\}$, compute the interval $[\underline{T_i},\overline{T_i}]$ such that $1 - \tanh^2(x_i)\in[\underline{T_i},\overline{T_i}]$ for all $x_i \in [\underline{x}_i, \overline{x}_i]$.
    \item Create an interval diagonal matrix $[\underline{D}, \overline{D}] = \text{diag}([\underline{T_1},\overline{T_1}],\dots,[\underline{T_5},\overline{T_5}])$.
    \item Compute the matrix product $W \cdot [\underline{D}, \overline{D}]$ using interval arithmetics: 
    $$(W \cdot [\underline{D}, \overline{D}])_{ij}=\sum_{k=1}^nW_{ik}\cdot[\underline{D}_{kj}, \overline{D}_{kj}]=W_{ij}\cdot[\underline{T}_j, \overline{T}_j]=\begin{cases}[W_{ij}\underline{T}_j, W_{ij}\overline{T}_j]\text{ if }W_{ij}\geq0\\ [W_{ij}\overline{T}_j, W_{ij}\underline{T}_j]\text{ otherwise}\end{cases}$$
    \item Add the constant term $\tau I_5$ to get $[\underline{J}_x, \overline{J}_x] = \tau I_5 + W \cdot [\underline{D}, \overline{D}]$
         $$J_x(x) = [\underline{J}_x, \overline{J}_x] = \begin{bmatrix}[J_{\text{lb}}(1,1), J_{\text{ub}}(1,1)] & [J_{\text{lb}}(1,2), J_{\text{ub}}(1,2)] & \cdots & [J_{\text{lb}}(1,5), J_{\text{ub}}(1,5)] \\ [J_{\text{lb}}(2,1), J_{\text{ub}}(2,1)] & [J_{\text{lb}}(2,2), J_{\text{ub}}(2,2)] & \cdots & [J_{\text{lb}}(2,5), J_{\text{ub}}(2,5)] \\ \vdots & \vdots & \ddots & \vdots \\ [J_{\text{lb}}(5,1), J_{\text{ub}}(5,1)] & [J_{\text{lb}}(5,2), J_{\text{ub}}(5,2)] & \cdots & [J_{\text{lb}}(5,5), J_{\text{ub}}(5,5)] \end{bmatrix} $$


\end{enumerate}

\section{Mixed Monotonicity}\label{apd:nODE Mixed Monotonicity}

In this appendix, we adapt the continuous-time mixed monotonicity and sampled-data mixed monotonicity methods from \citet{meyer2021interval}, which were originally established  for continuous-time dynamical systems, to neural ODE.

\subsection{Continuous-Time Mixed Monotonicity}\label{apd:CT mixed monotonicity}

The autonomous neural ODE \eqref{eq:nODE} is mixed monotone if there exists a decomposition function $g: \mathbb{R}^n \times \mathbb{R}^n \to \mathbb{R}^n$ such that for all $x, \hat{x} \in \mathbb{R}^n$, the following conditions hold:
\begin{itemize}
    \item $g$ is increasing in its first argument (off-diagonally):
    $$\forall i, j \in \{1, \dots, n\}, \, j \neq i: \quad \frac{\partial g_i}{\partial x_j}(x, \hat{x}) \geq 0,$$
    \item $g$ is decreasing in its second argument:
    $$\forall i, j \in \{1, \dots, n\}: \quad \frac{\partial g_i}{\partial \hat{x}_j}(x, \hat{x}) \leq 0,$$
    \item $f$ is embedded in the diagonal of $g$:$$g(x,x) = f(x),$$
\end{itemize}

This decomposition implies that the embedded dynamical system is evolving in $\mathbb{R}^{2n_x}$:
$$\begin{bmatrix} \dot{x} \\ \dot{\hat{x}} \end{bmatrix} = \begin{bmatrix} g(x, \hat{x}) \\ g(\hat{x}, x) \end{bmatrix} = h(x, \hat{x}),$$
is monotone with respect to the orthant $\mathbb{R}_{+}^{n_{x}} \times \mathbb{R}_{-}^{n_{x}}$ in its state space.
\\
\\
\textbf{Requirements and Limitations}: For applicability, there must exist a matrix $L_x \in \mathbb{R}^{n_x \times n_x}$ such that $J(x) + L_x$ is sign-stable over the considered time and state ranges (where $J(x) = \frac{\partial f}{\partial x}(x)$ is the Jacobian matrix). This means that each off-diagonal element of $J(x) + L_x$ maintains a constant sign (positive or negative) for all $x$ in the domain (e.g., the reachable tube estimate containing all trajectories over $[0, T]$). 

To construct $L_x$ while minimizing the conservatism of the over-approximation, \citet{meyer2021interval} recommends shifting each off-diagonal Jacobian element based on its interval bounds $[\underline{J}_{x_{ij}}, \overline{J}_{x_{ij}}]$. Namely, for each $i, j$ with $j \neq i$, the shifting value $y = L_{x_{ij}}$ is chosen to move the interval to the nearest sign-stable half-plane with minimal distance: 
$$ y= \begin{cases} 
\max(0, -\underline{J}_{x_{ij}}) & \text{if } |\underline{J}_{x_{ij}}| \leq |\overline{J}_{x_{ij}}|, \\
\min(0, -\overline{J}_{x_{ij}}) & \text{if } |\underline{J}_{x_{ij}}| > |\overline{J}_{x_{ij}}|.
\end{cases}$$
which handles the cases where the interval is already sign-stable (no shift, $y=0$) or crosses zero (shift by a smaller overhang). 

The decomposition function $g$ embeds the vector field such that its $i$-th component is:
$$g_i(x, \hat{x}) = f_i(\xi_i) + \sum_{j=1}^n |L_{x_{ij}}| (x_j - \hat{x}_j),$$
where $\xi_i \in \mathbb{R}^n$ is defined component wise as $\xi_{i j} = x_j$ if $L_{x_{ij}} \geq 0$, else $\xi_{i j} = \hat{x}_j$

The reachable set at time $T$ from initial interval $[\underline{x_0},\overline{x_0}]$ is over-approximated by simulating the embedded system only once from the initial condition $\begin{bmatrix} x(0) \\ \hat{x}(0) \end{bmatrix}=\begin{bmatrix} \underline{x_0} \\ \overline{x_0} \end{bmatrix}$: the $n_x$ first output variables of the embedded system represent the lower bound of this over-approximation; and the $n_x$ last output variables its upper bound. This method is computationally efficient for neural ODE, although conservatism may increase with larger $T$.

\begin{algorithm2e}
\caption{Continuous-Time Mixed Monotonicity Reachability for Neural ODE}
\label{alg:continuous-time-mm-nODE}
\DontPrintSemicolon
\KwIn{Neural ODE \eqref{eq:nODE}, initial set \([\underline{x}_0, \overline{x}_0] \subseteq \mathbb{R}^n\), time horizon \(T > 0\), \textbf{optional:} reachable tube estimate \([\underline{x}, \overline{x}] \subseteq \mathbb{R}^n\) containing \(\{ x(t) \mid t \in [0, T], x(0) \in [\underline{x}_0, \overline{x}_0] \}\)}
\KwOut{Interval bounds \([\underline{x}(T), \overline{x}(T)]\) for the reachable set at time \(T\)}

\Begin{
  \If{\([\underline{x}, \overline{x}]\) is not provided}{
    Compute Lipschitz constant \(L_f\) of \(f\)\;
    Set \([\underline{x}, \overline{x}] \leftarrow [\underline{x}_0 - L_f T \cdot \mathbf{1}, \overline{x}_0 + L_f T \cdot \mathbf{1}]\)
  }
  
  \textbf{Jacobian bounds:}\;
  Compute \([\underline{J}_x, \overline{J}_x] \subseteq \mathbb{R}^{n \times n}\) for the Jacobian matrix \(J_x = \frac{\partial f}{\partial x}\) over \(x \in [\underline{x}, \overline{x}]\)\;
  
  \textbf{Shifting matrix \(L_x\):}\;
  Set \(L_x \leftarrow 0_{n \times n}\)\;
  \For{\(i \leftarrow 1\) \KwTo \(n\)}{
    \For{\(j \leftarrow 1\) \KwTo \(n\), \(j \neq i\)}{
      Set \(L_{x_{ij}} \leftarrow y\), where:
      \[
      y = \begin{cases} 
      \max(0, -\underline{J}_{x_{ij}}) & \text{if } |\underline{J}_{x_{ij}}| \leq |\overline{J}_{x_{ij}}|, \\
      \min(0, -\overline{J}_{x_{ij}}) & \text{if } |\underline{J}_{x_{ij}}| > |\overline{J}_{x_{ij}}|.
      \end{cases}
      \]
    }
  }
  
  \textbf{Decomposition function \(g(x, \hat{x})\):}\;
  \For{\(i \leftarrow 1\) \KwTo \(n\)}{
    Set \(\xi_i \leftarrow 0_n\)\;
    \For{\(j \leftarrow 1\) \KwTo \(n\)}{
      \If{\(L_{x_{ij}} \geq 0\)}{
        Set \(\xi_i^j \leftarrow x_j\)\;
      }
      \Else{
        Set \(\xi_i^j \leftarrow \hat{x}_j\)\;
      }
    }
    Set \(g_i(x, \hat{x}) \leftarrow f_i(\xi_i) + \sum_{j=1}^n |L_{x_{ij}}| (x_j - \hat{x}_j)\)\;
  }
  
  \textbf{Auxiliary system:}\;
  Simulate the system \(\begin{bmatrix} \dot{x} \\ \dot{\hat{x}} \end{bmatrix} = \begin{bmatrix} g(x, \hat{x}) \\ g(\hat{x}, x) \end{bmatrix}\) from \(t = 0\) to \(T\) with initial conditions $\begin{bmatrix} \underline{x_0} \\ \overline{x_0} \end{bmatrix}$ to obtain an output corresponding to $\begin{bmatrix} \underline{x}(T) \\ \overline{x}(T) \end{bmatrix}$;

  \Return{Interval bounds \([\underline{x}(T), \overline{x}(T)]\)}
}
\end{algorithm2e}

\subsection{Sampled-Data Mixed Monotonicity}\label{apd:SD mixed monotonicity}

The main idea in this method is to model the continuous-time system as a discrete-time system by sampling at specific time points, with initial conditions $x(t_0) = x(0)$, the sampled-data version is defined as:
$$x^+ = \Phi(t_f; t_0, x(0)),$$
where $\Phi(t_f; t_0, x(0))$ is the solution of the ODE at time $t_f$ starting from $x(0)$ at $t_0$. This allows us to apply discrete-time reachability methods to the continuous-time system by treating the evolution from $t_0$ to $t_f$ as a single discrete step. Thus, the sampled-data approach can be directly applied by considering the solution map $\Phi(t_f; t_0, x(0))$, which for neural ODE \eqref{eq:nODE} is computed by integrating the ODE defined by the neural network, as it is an autonomous system.

This method relies on the sensitivity matrix:
$$S_x(t_f; t_0, x(0)) = \frac{\partial \Phi(t_f; t_0, x(0))}{\partial x(0)},$$
which describes how small changes in the initial state $x(0)$ affect the state at time $t_f$. Based on Assumption 5.1 from \citet{meyer2021interval}, there exists a matrix $L_x$ such that for all $i, j \in \{1, \dots, n\}$, either:
    $$S_{x_{ij}}(t_f; t_0, x_0) + L_{x_{ij}} \geq 0 \quad \forall x_0 \in [\underline{x_0}, \overline{x_0}],$$
    or
    $$S_{x_{ij}}(t_f; t_0, x_0) + L_{x_{ij}} \leq 0 \quad \forall x_0 \in [\underline{x_0}, \overline{x_0}],$$
which enforces a sign-stability condition similar to that used in discrete-time mixed monotonicity.

To construct $L_x$ while minimizing the conservatism of the over-approximation, \citet{meyer2021interval} recommends shifting each sensitivity element based on its interval bounds $[\underline{S}_{x_{ij}}, \overline{S}_{x_{ij}}]$. For each $i, j$, the shifting value $y = L_{x_{ij}}$ is chosen to move the interval to the nearest sign-stable half-plane with minimal distance:
$$ y= \begin{cases} 
\max(0, -\underline{S}_{x_{ij}}) & \text{if } |\underline{S}_{x_{ij}}| \leq |\overline{S}_{x_{ij}}|, \\
\min(0, -\overline{S}_{x_{ij}}) & \text{if } |\underline{S}_{x_{ij}}| > |\overline{S}_{x_{ij}}|.
\end{cases}$$
which handles the cases where the interval is already sign-stable or crosses zero.

The decomposition function $g$  embeds the flow map such that its $i$-th component is:
$$g_i(t_0, x, \hat{x}) = \Phi_i(t_f; t_0, \xi_i) + \sum_{j=1}^n |L_{x_{ij}}| (x_j - \hat{x}_j).$$

Finally, the reachable set at time $t_f$ is over-approximated by:
$$\left[ g(t_0, \underline{x}, \overline{x}), g(t_0, \overline{x}, \underline{x}) \right],$$
similar to the discrete-time mixed monotonicity method, where only two evaluations of the decomposition function are needed to compute the bounds. This approach is particularly advantageous for larger time horizons, as it avoids the accumulating conservatism of continuous-time methods.

\begin{algorithm2e}
\caption{Sampled-Data Mixed Monotonicity Reachability for Neural ODE}
\label{alg:sampled-data-nODE}
\DontPrintSemicolon
\KwIn{Neural ODE \eqref{eq:nODE}, initial set \([\underline{x}_0, \overline{x}_0] \subseteq \mathbb{R}^n\), time horizon \(T > 0\), \textbf{optional:} reachable tube estimate \([\underline{x}, \overline{x}] \subseteq \mathbb{R}^n\) containing \(\{ x(t) \mid t \in [0, T], x(0) \in [\underline{x}_0, \overline{x}_0] \}\)}
\KwOut{Interval bounds \([\underline{x}(T), \overline{x}(T)]\) for the reachable set over-approximation at time \(T\)}
\Begin{
  \If{\([\underline{x}, \overline{x}]\) is not provided}{
    Compute Lipschitz constant \(L_f\) of \(f\)\;
    Set \([\underline{x}, \overline{x}] \leftarrow [\underline{x}_0 - L_f T \cdot \mathbf{1}, \overline{x}_0 + L_f T \cdot \mathbf{1}]\)
  }

  \textbf{Jacobian bounds:}\;
  Compute \([\underline{J}_x, \overline{J}_x] \subseteq \mathbb{R}^{n \times n}\) for the Jacobian matrix \(J_x = \frac{\partial f}{\partial x}\) over \(x \in [\underline{x}, \overline{x}]\)\;

  \textbf{Sensitivity bounds:}\;
  Compute \([\underline{S}_x, \overline{S}_x] \subseteq \mathbb{R}^{n \times n}\) for the sensitivity matrix \(S_x(T; 0, x_0) = \frac{\partial \Phi(T; 0, x_0)}{\partial x_0}\) over \(x(0) \in [\underline{x}_0, \overline{x}_0]\)\;
  
  \textbf{Shifting matrix \(L_x\):}\;
  Set \(L_x \leftarrow 0_{n \times n}\)\;
  \For{\(i \leftarrow 1\) \KwTo \(n\)}{
    \For{\(j \leftarrow 1\) \KwTo \(n\), \(j \neq i\)}{
      Set \(L_{x_{ij}} \leftarrow y\), where:
      \[
      y = \begin{cases} 
      \max(0, -\underline{S}_{x_{ij}}) & \text{if } |\underline{S}_{x_{ij}}| \leq |\overline{S}_{x_{ij}}|, \\
      \min(0, -\overline{S}_{x_{ij}}) & \text{if } |\underline{S}_{x_{ij}}| > |\overline{S}_{x_{ij}}|.
      \end{cases}
      \]
    }
  }
  
  \textbf{Decomposition function \(g(x, \hat{x})\):}\;
  \For{\(i \leftarrow 1\) \KwTo \(n\)}{
    Set \(\xi_i \leftarrow 0_n\)\;
    \For{\(j \leftarrow 1\) \KwTo \(n\)}{
      \If{\(L_{x_{ij}} \geq 0\)}{
        Set \(\xi_i^j \leftarrow x_j\)\;
      }
      \Else{
        Set \(\xi_i^j \leftarrow \hat{x}_j\)\;
      }
    }
    Solve ODE \(\dot{x} = f(x)\) from \(t = 0\) to \(T\) with \(x(0) = \xi_i\) to obtain \(\Phi(T; 0, \xi_i)\)\;
    Set \(g_i(x, \hat{x}) \leftarrow \Phi_i(T; 0, \xi_i) + \sum_{j=1}^n |L_{x_{ij}}| (x_j - \hat{x}_j)\)\;
  }
  
  \textbf{Reachability bounds:}\;
  Compute \(g(\underline{x}_0, \overline{x}_0)\) and \(g(\overline{x}_0, \underline{x}_0)\)\;
  Set \(\underline{x}(T) \leftarrow g(\underline{x}_0, \overline{x}_0)\) and \(\overline{x}(T) \leftarrow g(\overline{x}_0, \underline{x}_0)\)\;
  
  \Return{Interval bounds \([\underline{x}(T), \overline{x}(T)]\)}
}
\end{algorithm2e}

\subsection{Boundary-Based Mixed Monotonicity Reachability Analysis for neural ODE}\label{apd:boundary-mm}

The boundary-based mixed monotonicity neural ODE reachability approach, outlined in Algorithm~\ref{alg:boundary-mm-nODE}, computes the over-approximation of the reachable set $\Omega(\mathcal{X}_{in})$ of the neural ODE \eqref{eq:nODE} from the boundaries of the initial input set $\mathcal{X}_{in}$ instead of the entire input set.

The algorithm begins by extracting the boundaries $\mathcal{B}$ of $\mathcal{X}_{in}$ (typically the $2 \times n$ facets of the hyperplanes as illustrated in \figureref{fig:spiral_CORA_with_Black_Points,fig:FPA_10_Boundary_points_TIRA}), and the output over-approximation set $\Omega(\mathcal{X}_{in})$ is initialized as empty.
For each of the extracted boundaries, the mixed monotonicity embedding is applied based on the mixed monotonicity method chosen from the two methods discussed in \sectionref{apd:CT mixed monotonicity,apd:SD mixed monotonicity}. 

Finally, the over-approximations $\Omega(\mathcal{B})$ from all boundaries are combined via a union, and we then take the interval hull of this union to ensure that it also contains the interior of the reachable set.

\begin{algorithm2e}
\caption{Boundary-Based Mixed Monotonicity Reachability for Neural ODE}
\label{alg:boundary-mm-nODE}
\DontPrintSemicolon
\KwIn{neural ODE \eqref{eq:nODE}, initial set \([\underline{x}_0, \overline{x}_0] \subseteq \mathbb{R}^n\), time horizon \(T > 0\), single-step mixed monotonicity}
\KwOut{Over-approximation \(\Omega(\mathcal{X}_{\text{in}})\) of the reachable set at time \(T\)}

\Begin{
  Set \(\mathcal{X}_{\text{in}} \leftarrow [\underline{x}_0, \overline{x}_0]\)\;
  Extract the boundaries \(\mathcal{B}\) of \(\mathcal{X}_{\text{in}}\)\;
  Initialize \(\Omega(\mathcal{X}_{\text{in}}) \leftarrow \emptyset\)\;
  
  \For{each boundary \(\mathcal{B}\) of \(\mathcal{X}_{\text{in}}\)}{
    Apply mixed monotonicity embedding to compute decomposition function \(g(x, \hat{x})\) based on Jacobian bounds\;
    Simulate the embedded system \(\begin{bmatrix} \dot{x} \\ \dot{\hat{x}} \end{bmatrix} = \begin{bmatrix} g(x, \hat{x}) \\ g(\hat{x}, x) \end{bmatrix}\) over \([0, T]\) with initial bounds from \(\mathcal{B}\)\;
    Obtain over-approximation \(\Omega(\mathcal{B})\)\;
    Set \(\Omega(\mathcal{X}_{\text{in}}) \leftarrow \Omega(\mathcal{X}_{\text{in}}) \cup \Omega(\mathcal{B})\)\;
  }
  Set \(\Omega(\mathcal{X}_{\text{in}})\) as the interval hull of itself\;
  \Return{Interval over-approximation \(\Omega(\mathcal{X}_{\text{in}})\)}
}
\end{algorithm2e}

\newpage
\section{Numerical Results}\label{apd:Detailed Numerical Results}

This appendix provides separate numerical results for each of the Spiral and FPA examples for CORA, NNV2.0, and TIRA toolboxes.
\tableref{tab:Spiral_results,tab:FPA_results} contain the tightness metrics that represent the ratio of the area of the computed reachable set over-approximations to the area spanned by the sampled successors in each 2D projection\footnote{The FPA example results are appearing over three subfigures, each representing the projection of two dimensions only as the FPA is a 5-dimensional example.}, as well as the computation times. The tightness metrics ratio is used to evaluate the conservatism of the over-approximation, where higher values indicate a more conservative (larger) over-approximation.

\subsection{Spiral}

\subsubsection{CORA}

\begin{figure}[htbp]
\floatconts
  {fig:spiral_CORA_with_Black_Points}
  {\caption{CORA Full Reachable Set OA vs. Boundaries only OA for Spiral at $t=1$}}
  {\includegraphics[width=0.65\linewidth]{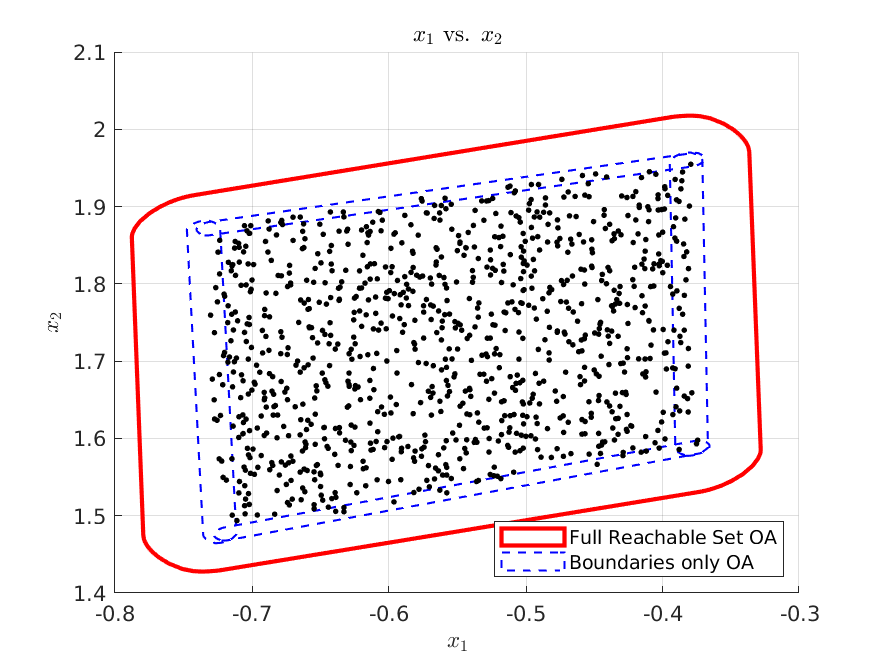}}
\end{figure}

\subsubsection{NNV2.0}
\textcolor{white}{SSS}
\begin{figure}[htbp]
\floatconts
  {fig:spiral_NNV2.0_with_Black_Points}
  {\caption{NNV2.0 Full Reachable Set OA for Spiral at $t=1$}}
  {\includegraphics[width=0.65\linewidth]{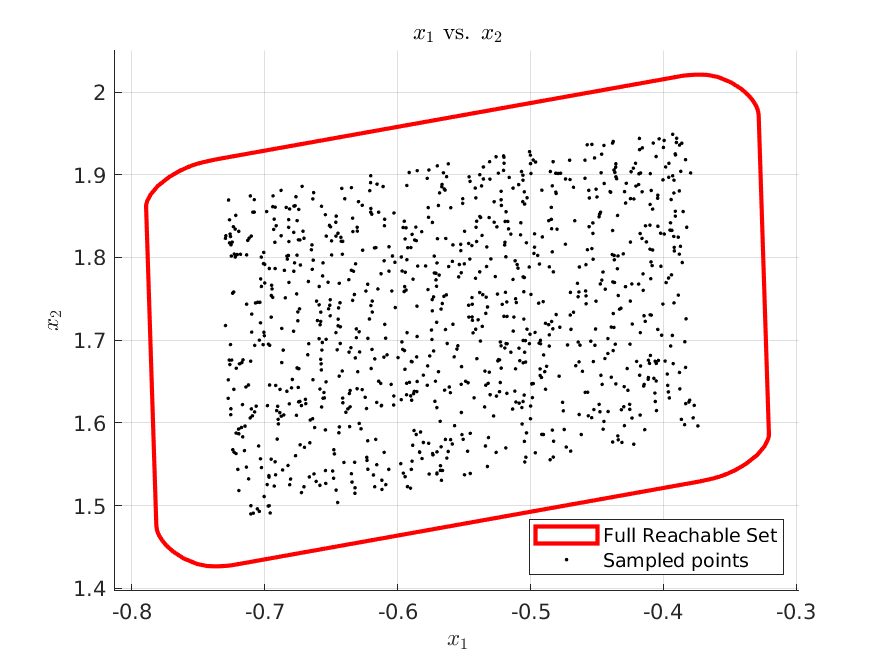}}
\end{figure}

\subsubsection{TIRA}

\begin{figure}[htbp]
\floatconts
  {fig:spiral_TIRA_with_legend}
  {\caption{TIRA single-step vs. incremental vs. boundary for Spiral at $t=1$}}
  {\includegraphics[width=0.7\linewidth]{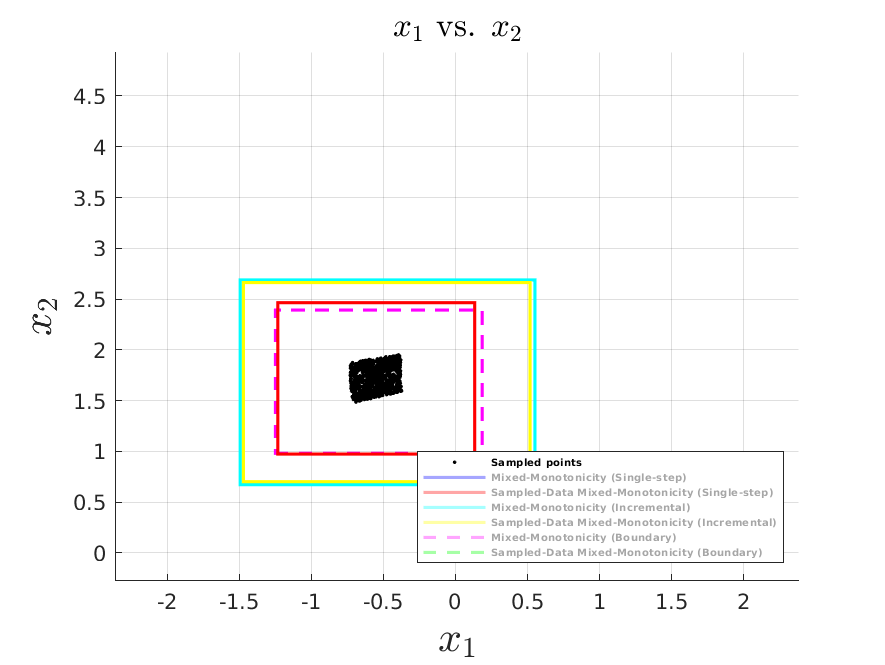}}
\end{figure}

\begin{table}[htbp]
\floatconts
  {tab:Spiral_results}
  {\caption{Spiral Numerical results at $t=1$ with CORA, NNV2.0, and TIRA}}%
  {%
    \begin{tabular}{|l|l|l|}
    \hline
    \abovestrut{2.2ex}\bfseries Methods & $\boldsymbol{x_1 - x_2}$ & \bfseries Time(sec.)\\\hline
    \abovestrut{2.2ex} CORA Full Reachable Set& 1.61 & 19.64 \\
    CORA Boundaries only & 1.15 &  70.83\\
    NNV2.0 Full Reachable Set& 1.71 & 17.25 \\
    TIRA (single-step) Mixed-Monotonicity & 24.59 & 0.66 \\
    TIRA (single-step) Sampled-Data Mixed-Monotonicity & 12.14 & 0.95 \\
    TIRA (incremental) Mixed-Monotonicity & 24.59 & 63.13 \\
    TIRA (incremental) Sampled-Data Mixed-Monotonicity & 23.24 & 111.16 \\
    TIRA (Boundary) Mixed-Monotonicity & 12.05 & 2.84 \\
    \belowstrut{0.2ex}TIRA (Boundary) Sampled-Data Mixed-Monotonicity & 12.14 & 4.35 \\\hline
    \end{tabular}
  }
\end{table}

\newpage
\subsection{FPA}

\subsubsection{CORA}

\begin{figure}[htbp]
\floatconts
  {fig:FPA_CORA_Full_vs_Boundaries_t_2}
  {\caption{ CORA Full Reachable Set OA vs. Boundaries only OA for FPA at $t = 2$}}
  {%
    \subfigure[$x_1 - x_2$]{\label{fig:single-step_x1-x2}%
      \includegraphics[width=0.6\linewidth]{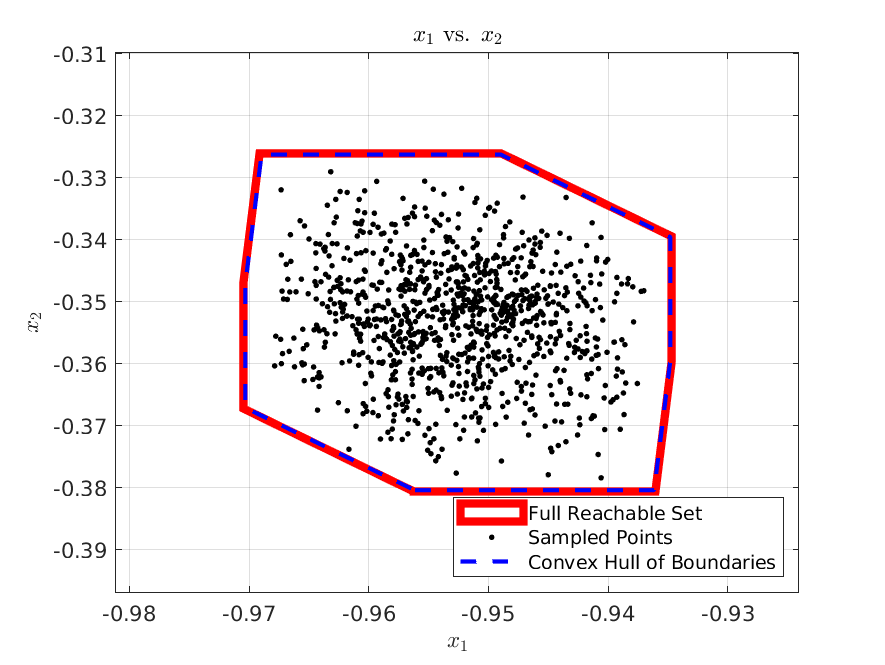}}%
    \subfigure[$x_3 - x_4$]{\label{fig:single-step_x3-x4}%
      \includegraphics[width=0.6\linewidth]{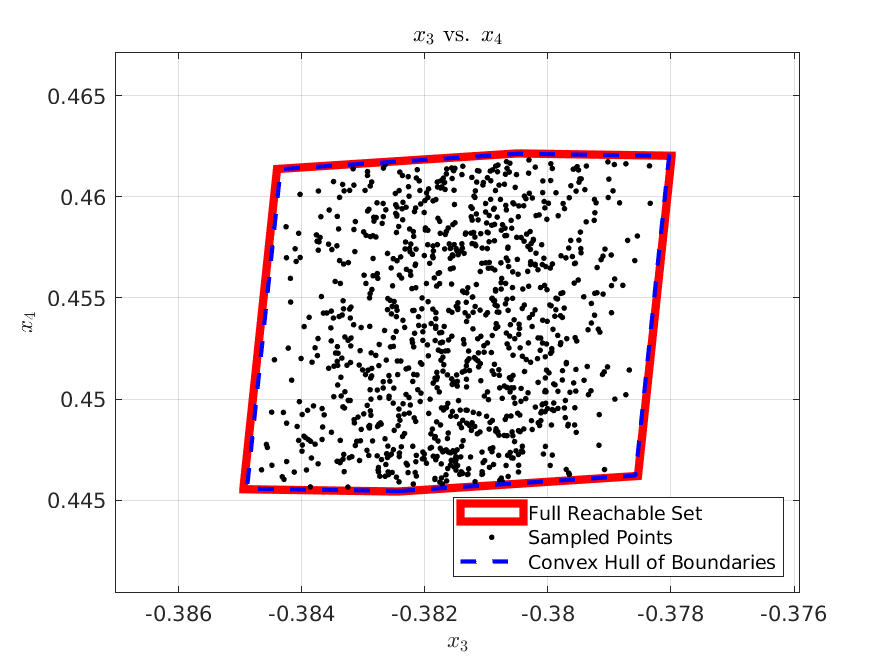}}
    \subfigure[$x_4 - x_5$]{\label{fig:single-step_x4-x5}%
      \includegraphics[width=0.6\linewidth]{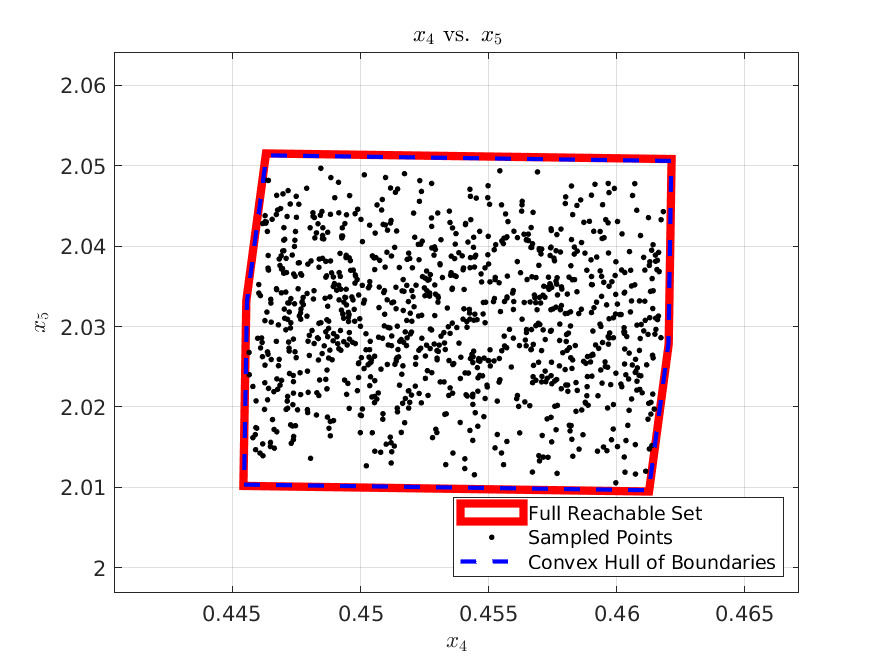}}
  }
\end{figure}

\clearpage
\subsubsection{NNV2.0}

\begin{figure}[htbp]
\floatconts
  {fig:FPA_NNV2.0_Full_OA_t_2}
  {\caption{ NNV2.0 Full Reachable Set OA at $t = 2$}}
  {%
    \subfigure[$x_1 - x_2$]{\label{fig:FPA_NNV2.0_x1-x2}%
      \includegraphics[width=0.6\linewidth]{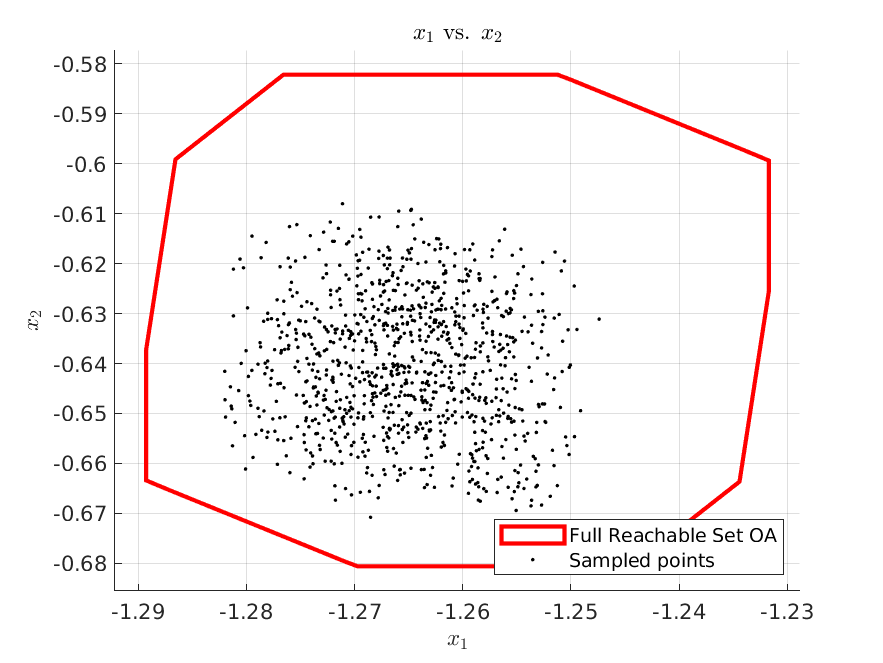}}%
    \subfigure[$x_3 - x_4$]{\label{fig:FPA_NNV2.0_x3-x4}%
      \includegraphics[width=0.6\linewidth]{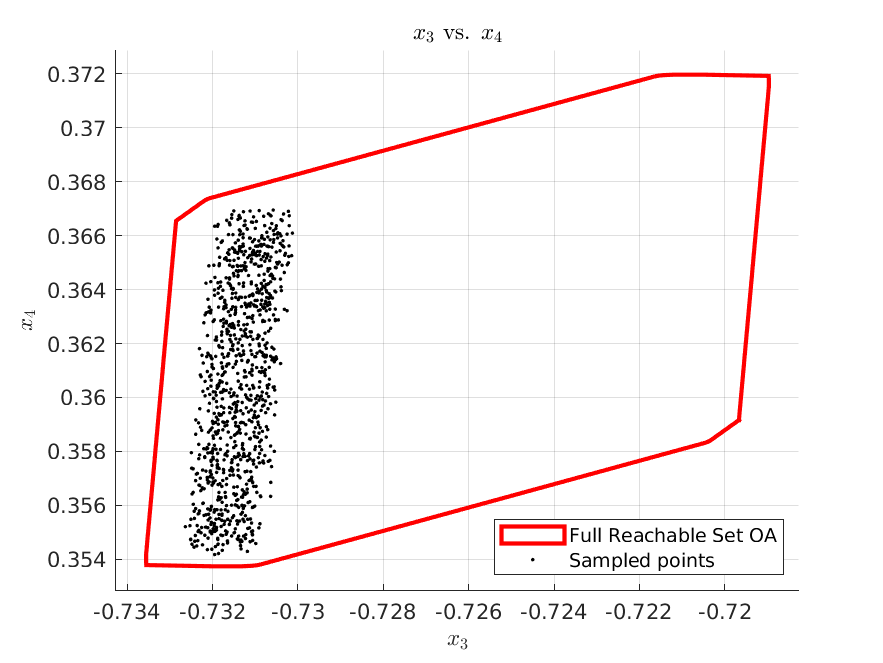}}
    \subfigure[$x_4 - x_5$]{\label{fig:FPA_NNV2.0_x4-x5}%
      \includegraphics[width=0.6\linewidth]{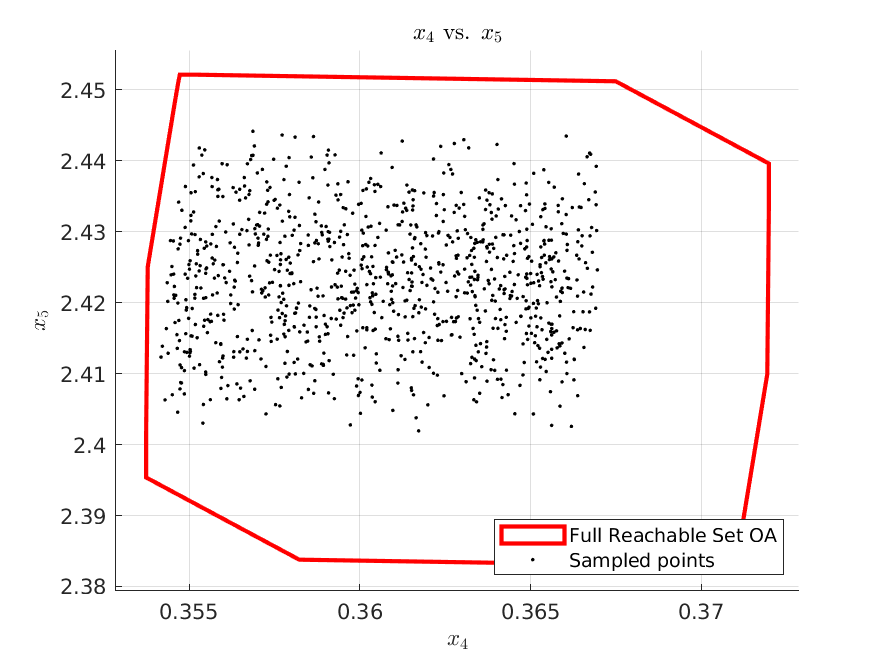}}
  }
\end{figure}

\clearpage
\subsubsection{TIRA}

\begin{figure}[htbp]
\floatconts
  {fig:FPA_TIRA_with_legend}
  {\caption{TIRA single-step vs. incremental vs. boundary for FPA at $t=2$}}
  {%
    \subfigure[$x_1 - x_2$]{\label{fig:TIRA_FPA_x1-x2}%
      \includegraphics[width=0.6\linewidth]{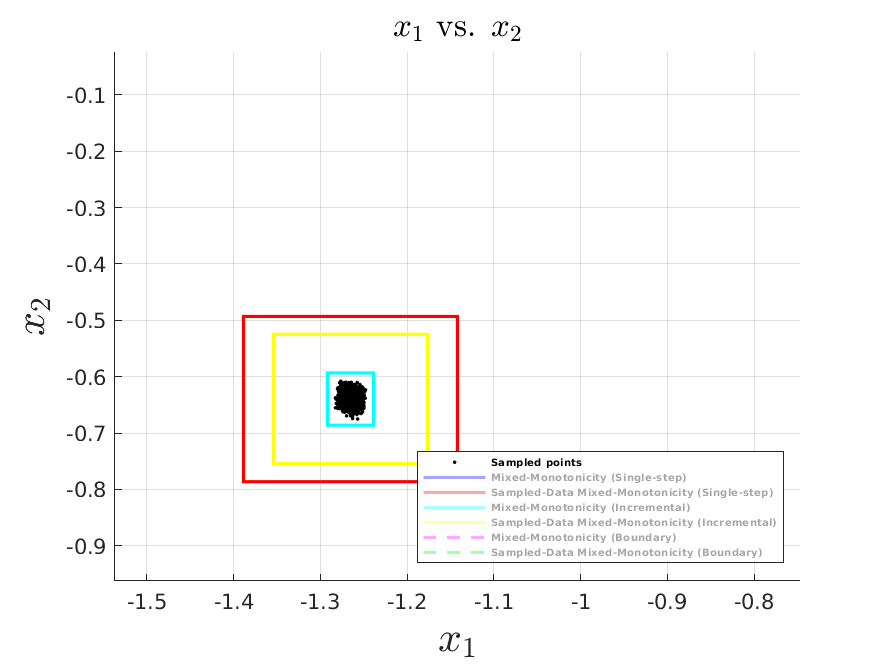}}%
    \subfigure[$x_3 - x_4$]{\label{fig:TIRA_FPA_x3-x4}%
      \includegraphics[width=0.6\linewidth]{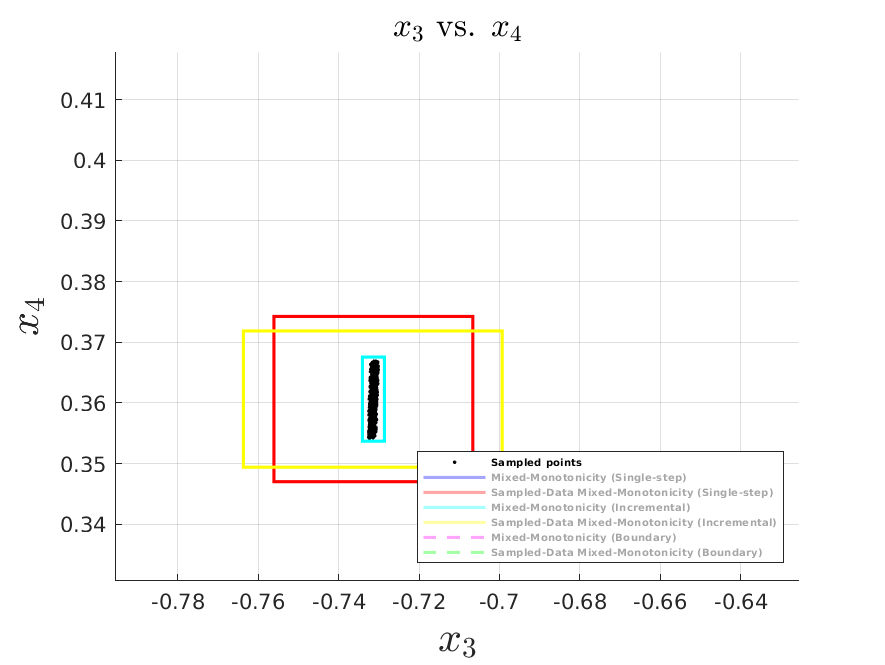}}
    \subfigure[$x_4 - x_5$]{\label{fig:TIRA_FPA_x4-x5}%
      \includegraphics[width=0.6\linewidth]{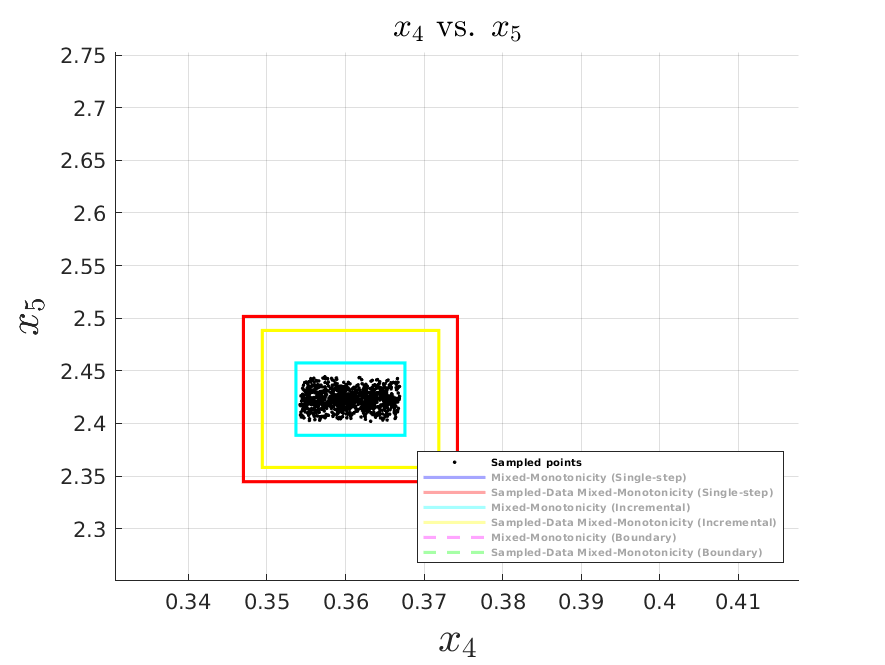}}
  }
\end{figure}

\begin{table}[htbp]
\floatconts
  {tab:FPA_results}
  {\caption{FPA Numerical results at $t=2$ with CORA, NNV2.0, and TIRA}}%
  {%
    \begin{tabular}{|l|l|l|l|l|}
    \hline
    \abovestrut{2.2ex}\bfseries Methods & $\boldsymbol{x_1 - x_2}$ & $\boldsymbol{x_3 - x_4}$ & $\boldsymbol{x_4 - x_5}$ & \bfseries Time(sec.)\\\hline
    \abovestrut{2.2ex}CORA Full Reachable Set & 1.33 & 1.11 & 1.13 & 13.22\\
    CORA Boundaries only & 1.18 & 0.99 & 1.08 &  109.1\\
    NNV2.0 Full Reachable Set & 2.52 & 8.74 & 2.43 & 11.98\\
    TIRA (single-step) Mixed-Monotonicity & 2.29 & 2.30 & 1.79 & 0.83 \\
    TIRA (single-step) Sampled-Data Mixed-Monotonicity & 33.57 & 40.67 & 8.05 & 1.34 \\
    TIRA (incremental) Mixed-Monotonicity & 2.29  & 2.30 & 1.79 & 25.41 \\
    TIRA (incremental) Sampled-Data Mixed-Monotonicity & 18.92 & 43.64 & 5.50 & 48.06 \\
    TIRA (Boundary) Mixed-Monotonicity & 2.29 & 2.30 & 1.79 & 7.06\\
    \belowstrut{0.2ex}TIRA (Boundary) Sampled-Data Mixed-Monotonicity & 33.57 & 40.67 & 8.05 & 12.76\\\hline
    \end{tabular}
  }
\end{table}

\begin{figure}[htbp]
\floatconts
  {fig:FPA_10_Boundary_points_TIRA}
  {\caption{FPA 10 Boundaries}}
  {%
    \subfigure[$x_1 - x_2$]{\label{fig:boundaries_x1-x2}%
      \includegraphics[width=0.6\linewidth]{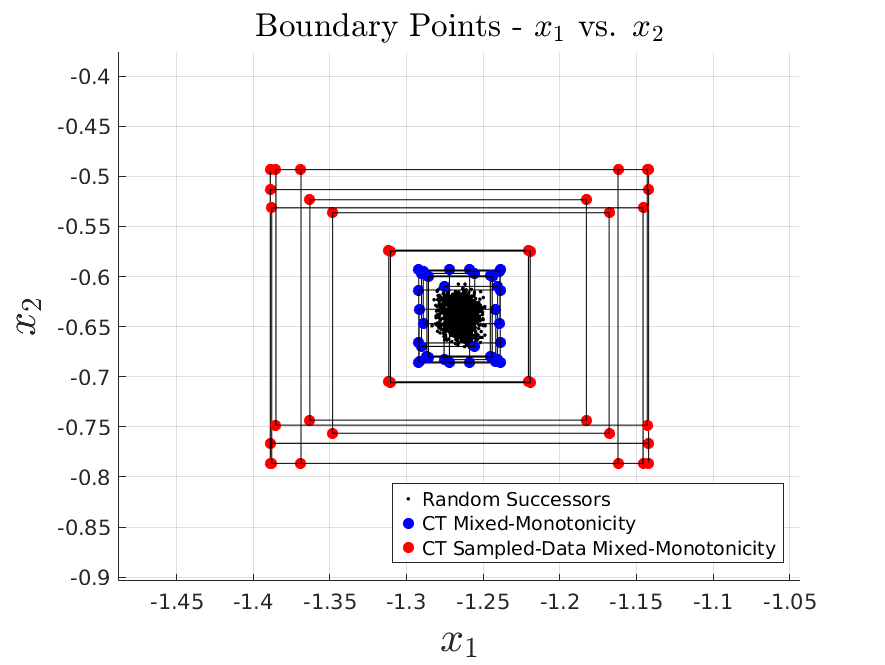}}%
    \subfigure[$x_3 - x_4$]{\label{fig:boundaries_x3-x4}%
      \includegraphics[width=0.6\linewidth]{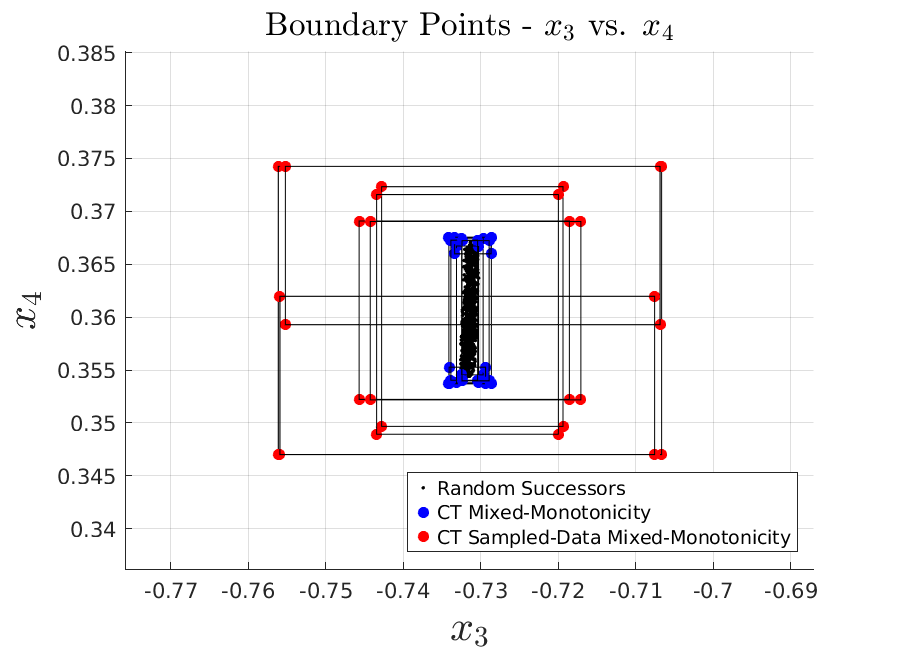}}
    \subfigure[$x_4 - x_5$]{\label{fig:boundaries_x4-x5}%
      \includegraphics[width=0.6\linewidth]{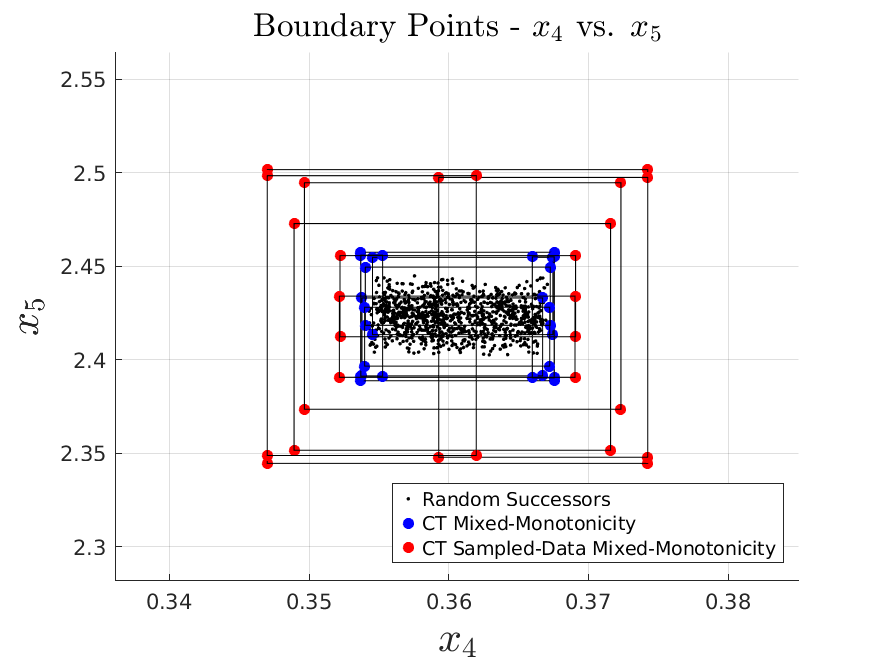}}
}
\end{figure}

\end{document}